# The impact of the fourth SM family on the Higgs observability at the LHC

E. Arik[1] , S.A. Cetin[2] and S. Sultansoy[3,4]
[1] Bogazici University, Istanbul, Turkey; [2] Dogus University, Istanbul, Turkey,
[3] TOBB University of Economics and Technology, Ankara, Turkey,
[4] Institute of Physics, Academy of Sciences, Baku, Azerbaijan

### Abstract

It is shown that if the fourth SM fermion family exists then the Higgs boson could be observed at the LHC with an integrated luminosity of few fb⁻¹. The Higgs discovery potential for different channels is discussed in the presence of the fourth SM family.

It is known that the Standard Model (SM) does not predict the number of fermion families, N. The only restriction comes from the asymptotic freedom of QCD which requires the number of quarks to be less than 17 and therefore, the number of SM families to be N ≤ 8. Before 1990's, many authors published articles related to the extra SM families and their phenomenological consequences. In early 1990's, the LEP data yielded N = 3 where the neutral lepton mass for each family is less than half the mass of the Z boson [1]. Most of the time, this result is interpreted as the exact value of N, since one assumes that the neutrinos must have very small masses. If we disregard this incorrect assumption, the LEP data does not exclude the existence of extra SM families with heavy neutrinos. Meanwhile, few papers arguing the existence of the fourth SM family were published [2-4]. These arguments were based on the 'flavor democracy' hypothesis (see [5] and references therein).

The fourth SM family leads to an essential increase in the Higgs boson production cross section via gluon fusion at the hadron colliders [6-11]. This motivated CDF and D0 collaborations to intensify their search for the Higgs boson in the gg → H → WW channel [12, 13] for which they had no sensitivity in the 3 SM family case.

In this note we consider the impact of the infinitely heavy fourth SM family in all the observable decay modes of the Higgs boson [14] at the LHC. Figure 1 shows the sensitivities of ATLAS and CMS for the discovery of the Higgs boson in 3 SM family case for 30 fb⁻¹ integrated luminosity [15]. The change in the Higgs discovery potential of ATLAS in the case of 4 SM families can be estimated by rescaling Figure 1. The results are presented in Figure 2.

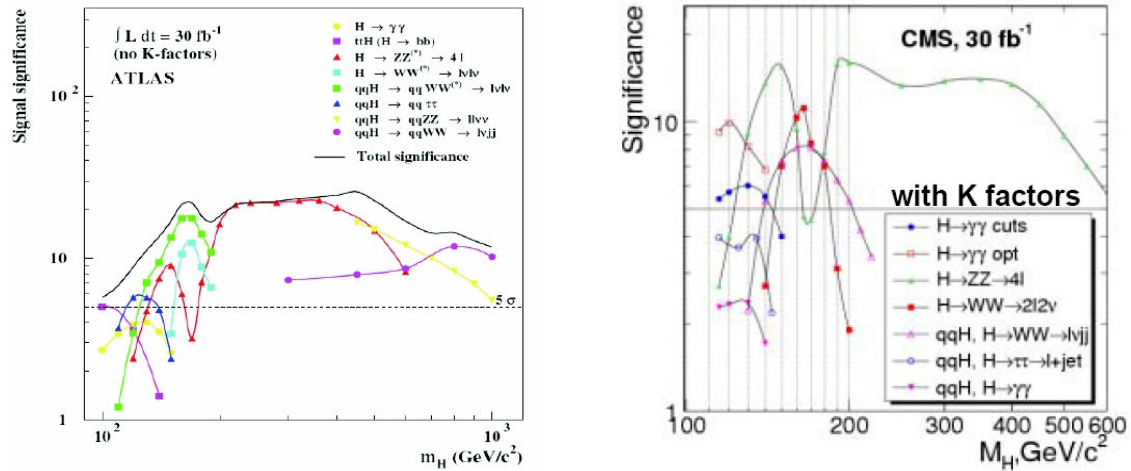

**Fig. 1:** ATLAS (left) and CMS (right) sensitivities for the discovery of the SM Higgs boson in the 3 SM family case[15].

In reference [9] it has already been shown that the Higgs boson can be observed at the LHC via the golden mode even with a few fb$^{-1}$ integrated luminosity if the fourth SM family exists. In addition to this, we see from Figure 2 that the WW mode is also well improved. Most of the decay modes that are studied for ATLAS in Figure 1 are through the associated production of Higgs, hence they are not influenced by the existence of a new family except for some decrease in their branching ratios due to the enlargement of the total Higgs decay width. It is only the $\gamma\gamma$, ZZ and WW channels that are effected, where for the case of the $\gamma\gamma$ mode, the enhancement in the gg→H production is almost compensated by the decrease in its decay width because of the negative contribution.

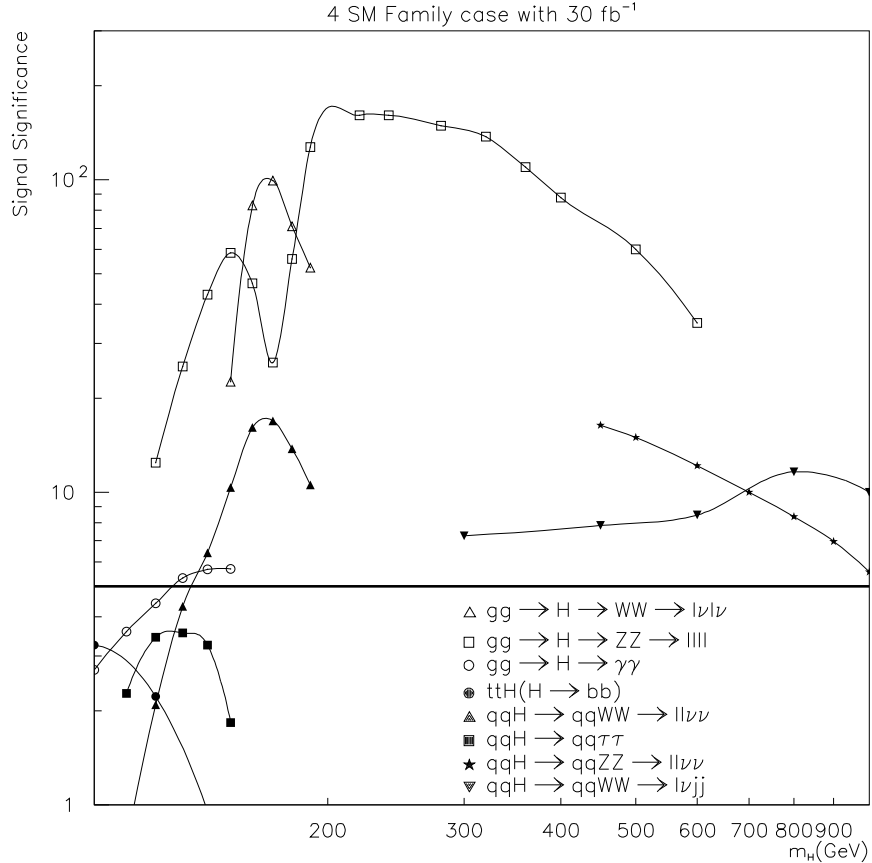

**Fig. 2:** ATLAS sensitivity for the discovery of the SM Higgs boson in the 4 SM family case with an integrated luminosity of 30 fb$^{-1}$.

The comparison of Figure 1 and 2 for ATLAS gives us the hint that one could get significant results before reaching 30 fb$^{-1}$ integrated luminosity. This is well demonstrated in Figure 3a where signal significances for 10 fb$^{-1}$ are given. It is seen that both the golden mode and WW channel, as well some qqH channels still give significance of at least 5, however for those qqH channels the number of events approach the critical low limit where one has to use Poisson statistics. This is why when we show the case of 3 fb$^{-1}$ in Figure 3b, we omit them not for their low significance but for the unrealistic number of events for discovery in those channels at such an integrated luminosity. Even at an integrated luminosity of 1 fb$^{-1}$ (Figure 3c), the ZZ and WW modes are promising at levels higher than 5σ for almost all of the higss mass region.

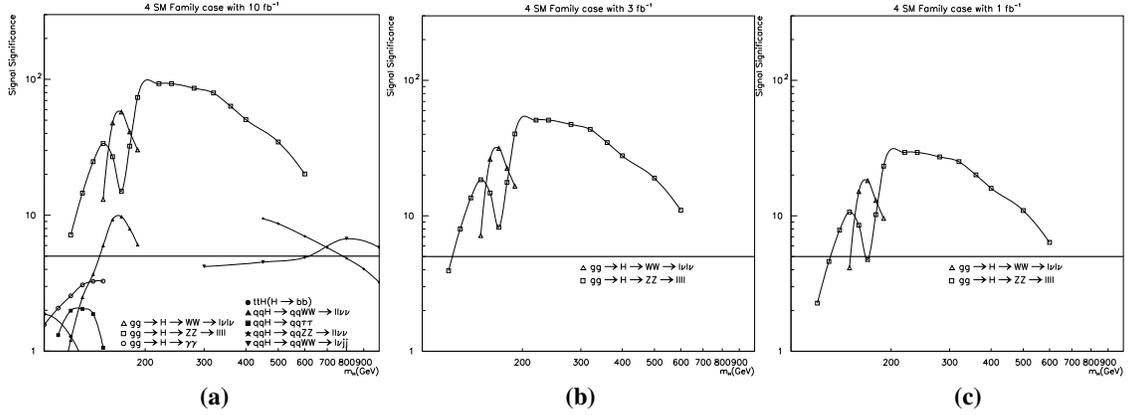

**Fig. 3:** ATLAS sensitivity for the discovery of the SM Higgs boson in the 4 SM family case with an integrated luminosity of  **(a)** 10 fb$^{-1}$, **(b)** 3 fb$^{-1}$ and **(c)** 1 fb$^{-1}$ .

It can be concluded that, even with 1 fb$^{-1}$, the WW mode will be an important channel for the discovery of the Higgs boson in the region 150-200 GeV. Such a discovery will assure the existence of the 4$^{th}$ family, whereas the golden mode will cover mass region until ~600 GeV with the same level of integrated luminosity.

Flavor democracy with four SM families explains the observed quark and charged-lepton mass spectrum (as well as quark and leptonic CKM mixings) in a most natural way being consistent with experimental data [16]. Possibly the TEVATRON or most probably the LHC data will yield the final confirmation of the fourth SM family within few years, owing to the enhancement of Higgs boson production due to the fourth family quarks. The LHC will be able to observe the direct production of the fourth family quarks too.

### Acknowledgements

S. Sultansoy would like to thank P. Jenni for the provided support during his visit to CERN. E. Arik and S. A. Cetin acknowledge the support from Turkish Atomic Energy Authorithy.